\documentclass[11pt]{article}
\usepackage{moriond_nkw,epsfig}
\usepackage{amssymb} 

\newcommand{\Mz}{\ensuremath{M_{\mathrm{Z}}}}
\newcommand{\Mw}{\ensuremath{M_{\mathrm{W}}}}
\newcommand{\Gw}{\ensuremath{\Gamma_{\mathrm{W}}}}
\newcommand{\WW}{\ensuremath{\mathrm{W}^+\mathrm{W}^-}}
\newcommand{\epem}{\ensuremath{\mathrm{e}^+\mathrm{e}^-}}
\newcommand{\Zqq}{\ensuremath{\Zz/\gamma\rightarrow q\overline{q}}}
\newcommand{\qq}{\ensuremath{\mathrm{q\overline{q}^\prime}}}
\newcommand{\lv}{\ensuremath{\ell\overline{\nu}_{\ell}}}
\newcommand{\tv}{\ensuremath{\tau{\nu}_{\tau}}}
\newcommand{\WWqqtv}{\ensuremath{\WW\rightarrow\qq\tv}}
\newcommand{\WWqqlv}{\ensuremath{\WW\rightarrow\qq\lv}}
\newcommand{\WWqqqq}{\ensuremath{\WW\rightarrow\qq\qq}}
\newcommand{\lpv}{\ensuremath{\ell^+ \nu_{\ell}}}
\newcommand{\lmv}{\ensuremath{{\ell^{\prime}}^-\overline{\nu}_{\ell^{\prime}}}}
\newcommand{\WWlvlv}{\ensuremath{\WW\rightarrow\lpv\lmv}}
\newcommand{\qqqq}{\ensuremath{\qq\qq}}
\newcommand{\qqlv}{\ensuremath{\qq\lv}}
\newcommand{\Dnch}{\ensuremath{\Delta\langle n_{\mathrm{ch}}\rangle}}
\newcommand{\nQQQQ}{\ensuremath{\langle n_{\mathrm{ch}}^{\mathrm{4q}}\rangle}}
\newcommand{\nQQLV}{\ensuremath{\langle n_{\mathrm{ch}}^{\mathrm{qq}\ell\nu}\rangle}}
\newcommand{\Aleph}{\mbox{A{\sc leph}}}
\newcommand{\Delphi}{\mbox{D{\sc elphi}}}
\newcommand{\Lthree}{\mbox{L{\sc 3}}}
\newcommand{\Opal}{\mbox{O{\sc pal}}}
\newcommand{\Lep}{\mbox{L{\sc ep}}}
\newcommand{\Lepone}{\mbox{L{\sc ep1}}}
\newcommand{\Leptwo}{\mbox{L{\sc ep2}}}
\newcommand{\roots}{\ensuremath{\sqrt{s}}}

\newcommand{\kt}{\ensuremath{k_{\perp}}}
\newcommand{\CC}{\mbox{{\sc CC03}}}
\newcommand{\Zz}{\ensuremath{{\mathrm{Z}^0}}}
\newcommand{\Zgamma}{\ensuremath{\Zz/\gamma}}
\newcommand{\Koralw}{\mbox{K{\sc oralw}}}
\newcommand{\Excalibur}{\mbox{E{\sc xcalibur}}}
\newcommand{\Jetset}{\mbox{J{\sc etset}}}
\newcommand{\Herwig}{\mbox{H{\sc erwig}}}
\newcommand{\Wp}{\ensuremath{\mathrm{W}^+}}
\newcommand{\Wm}{\ensuremath{\mathrm{W}^-}}
\newcommand{\pipip}{\ensuremath{\pi^+\pi^+}}
\newcommand{\pipim}{\ensuremath{\pi^-\pi^-}}
\newcommand{\pipid}{\ensuremath{\pi^+\pi^-}}
\newcommand{\pb}[1]{\ensuremath{\mathbf{p}_{#1}}}

\def\as{\ensuremath{\alpha_{s}}}
\def\etal{\mbox{{\it et al.}}}

\def\eg{\mbox{{\it e.g.}}}

\newcommand{\plb}[3]  {Phys.\ Lett.\ \textbf{B#1} (#2) #3}
\newcommand{\zpc}[3]  {Z.\ Phys.\ \textbf{C#1} (#2) #3}
\newcommand{\epc}[3]  {Eur.\ Phys.\ J.\ \textbf{C#1} (#2) #3}
\newcommand{\prl}[3]  {Phys.\ Rev.\ Lett.\ \textbf{#1} (#2) #3}

\begin{document}
\vspace*{4cm}
\title{W MASS AND \WW\ FINAL STATE INTERACTIONS}

\author{NIGEL K. WATSON}

\address{School of Physics and Astronomy, University of Birmingham, \\
         Edgbaston, Birmingham, B15 2TT, Great Britain}

\maketitle\abstracts{Precise measurements of the mass and width of the
 W boson are carried out in \epem\ collisions at \Lep, by kinematic
 reconstruction of the invariant mass distributions of \WWqqlv\ and
 \WWqqqq\ candidate events.  The most recent combination of such
 results from \Aleph, \Delphi, \Lthree\ and \Opal\ uses approximately 82\% of
 the final \Leptwo\ integrated luminosity and is preliminary.  The mass of
 the W boson so determined is $\Mw=80.446 \pm0.026\pm0.030$~GeV, while
 the corresponding direct measurement of the W boson width gives $\Gw
 = 2.148\pm0.071\pm0.063$~GeV.  These measurements are subject to
 sizeable systematic uncertainties from the QCD phenomena of
 Bose-Einstein correlations and colour reconnection.  Recent
 substantial progress in both of these areas is reported.}

\section{W Mass and Width}
The \Lep\ \epem\ collider at CERN has provided an ideal environment
for the study of the properties of the gauge bosons of the Standard
Model of electroweak interactions.  Since 1996, it operated at
centre-of-mass energies above the \WW\ production threshold (\Leptwo),
allowing direct measurements of the W boson mass, \Mw.  When combined
with the direct measurements of the top quark mass at the Tevatron,
these allow further constraints to be set on the mass of the Higgs
boson via electroweak radiative corrections.  Comparison between the
direct measurements of the mass of the W boson and the value
determined indirectly from data recorded at $\roots\approx\Mz$
provides an important test of the self-consistency of the Standard
Model.  The direct measurement of $\Gw$ further tests the consistency
of the Standard Model.

The measurement of \Mw\ and \Gw\ is divided into three stages:
selection of \WW\ events, event-by-event mass reconstruction, and
determination of \Mw\ itself.

\subsection{\WW\ Final States}
\WW\ events are divided into three final states.  \WWqqqq\ events
comprise $45\%$ of the total \WW\ cross-section and are characterised
by four energetic jets of hadrons with little or no missing energy.
Semi-leptonic \WWqqlv\ decays comprise $44\%$ of the total \WW\
cross-section and are characterised by two distinct hadronic jets, a
high-momentum lepton and missing momentum due to the prompt neutrino
from the leptonic W decay.  The signature for the \WWqqtv\ channel is
similar, with the exception that the $\tau$ lepton is identified as an
isolated, low-multiplicity jet typically consisting of one or three
tracks.  The \WWlvlv\ channel, with at least two unobserved neutrinos
and a relatively low branching fraction, has limited \Mw\ sensitivity
and is not discussed herein.

\subsection{Invariant Mass Reconstruction}

 The clean environment at \Lep\ allows a complete kinematic
 reconstruction of the invariant mass on an event-by-event basis.
 Hadrons are grouped together into jets using clustering algorithms
 such as \kt.  In \WWqqlv\ events, charged leptons are identified and
 neutrinos are inferred from the missing energy and momentum.  The
 invariant masses of the two W bosons can be determined directly from
 the reconstructed momenta of observed decay products. Experimentally,
 the limiting factor in the mass resolution is the uncertainty in the
 jet energy measurement, which is poor in contrast to the measured jet
 directions.  As the centre-of-mass energy is well known, the mass
 resolution can be improved significantly (factor $\sim$2--3) by use
 of a constrained kinematic fit imposing the four constraints of
 energy and momentum conservation (4-C fit).  Small additional gains
 are possible by imposing the additional constraint that the masses of
 the two W bosons are equal in each event (5-C fit), giving a single
 mass measurement per event.

 For \WWqqlv\ events, the number of effective constraints is reduced
 to 2 (1) for a 5-C (4-C) fit due to the three missing degrees of
 freedom corresponding to the unmeasured neutrino momentum.  For
 \WWqqtv\ events, most of the mass information is given by the
 hadronically decaying W.  A frequent assumption in constructing
 kinematic fits for these events is that the true $\tau$ direction
 coincides with its observed decay products, while the $\tau$ energy
 is unknown, removing a further constraint.

 In \WWqqqq\ events where four jets are reconstructed, there are three
 possible pairwise combinations.  In some analyses, a single preferred
 combination is selected on the basis of information such as kinematic
 fit probabilities, jet-jet angles or the \CC\footnote{
 Doubly-resonant \WW\ production diagrams, {\em i.e.} $t$-channel
 $\nu_{\mathrm{e}}$ exchange and $s$-channel \Zgamma\ exchange.}\
 matrix elements.  In other analyses, all combinations are used with
 different weights assigned to each, or two combinations are used with
 equal weight.  In addition, as quarks may radiate energetic gluons
 leading to a distinct five-jet topology, several analyses separate
 events into four- and five-jet categories to be treated separately,
 leading to an overall improved mass resolution.

\subsection{\Mw\ Determination}
 There are three main methods whereby \Mw\ is determined.  The most
 widely used (\Aleph, \Lthree, \Opal) involves reweighting Monte Carlo
 events including detector simulation to an arbitrary value of \Mw\
 using the ratio of 4-fermion or \CC\ matrix elements. A likelihood
 fit to the reweighted mass spectra determines the value of \Mw\ which
 best resembles the data.  The fit can be either a 1-dimensional fit to the
 5-C fit mass, possibly performed in several exclusive regions of
 reconstructed mass error, or a 2-dimensional fit to the 4-C fit
 masses.  There is an implicit MC correction, e.g. for effects such as
 initial state photon radiation (ISR) and detector effects.

 A second method (\Delphi, \Opal) constructs an event likelihood from
 the convolution of a Breit-Wigner with a radiator function (to
 account for ISR) and a resolution function (to account for detector
 effects).  The probability of an event being \WW\ is also included in
 the likelihood.  These `convolution' based methods make more use of
 the information per event and so might be expected to give a more
 precise \Mw\ determination.  An explicit bias correction has to be
 applied to the fitted \Mw, based on events including detector
 simulation.

 The third method (\Opal) fits an analytic function (asymmetric
 Breit-Wigner) to the reconstructed mass spectra.  Similarly, an
 explicit bias correction is made.  All methods yield comparable
 precision on the measured \Mw.

\subsection{Systematic Effects}
\begin{table}[htbp]
  \begin{center}
    \begin{tabular}{|l|c|c|c|} \hline
              &\multicolumn{3}{|c|}{ Uncertainty on \Mw\ (MeV) }   \\
               Source            &\qqlv&\qqqq  &Combined \\ \hline
      ISR/FSR                    &   8 &   8   &   7  \\
      Hadronisation              &  19 &  17   &  18  \\
      Detector                   &  11 &   8   &  10  \\
      Beam Energy                &  17 &  17   &  17  \\
      Colour Reconnection        & $-$ &  40   &  11  \\ 
      Bose-Einstein Correlations & $-$ &  25   &   7  \\
      Other                      &   4 &   5   &   3  \\ \hline
      Total systematic           &  29 &  54   &  30  \\ 
      Statistical                &  33 &  31   &  26  \\ \hline
      Total                      &  44 &  63   &  40  \\ \hline
    \end{tabular}
  \end{center}
  \caption{Summary of the systematic uncertainties for the
    combined \Lep\ fit results.}
  \label{tab:sys}
\end{table}
 The statistical uncertainty on the combined \Lep\ \Mw\ measurements
 is 26~MeV, therefore a clear understanding of systematic
 uncertainties is vitally important.  Systematics may be correlated
 between combinations of experiments, channels and years of data
 taking.  A summary of the uncertainties, without giving the detailed
 decomposition into the various correlated sources, is given in
 Table~\ref{tab:sys}.  A consequence of the large systematics from
 Bose-Einstein Correlations (BEC) and Colour Reconnection (CR) is that
 the \WWqqqq\ channel has a reduced contribution to the overall \Mw,
 with a weight of 0.27.  In the case where both channels had equal
 weight, the statistical uncertainty on \Mw\ would be reduced by
 $\sim15$\%.

 The most significant systematics are those which are correlated
 between experiments, as described below.  Photonic radiative
 corrections (ISR, FSR) have been estimated originally by comparing
 different ISR models, e.g. as in \Koralw\ and \Excalibur.  More
 recently, \Koralw\ events have been reweighted to correspond to an
 ${\cal{O}}(\alpha^2)$ or ${\cal{O}}(\alpha)$ treatment of ISR using
 the matrix elements calculated inside the model (the default
 calculation is at ${\cal{O}}(\alpha^3)$).  However, these
 calculations are incomplete at ${\cal{O}}(\alpha)$ (no ISR/FSR
 interference, or direct $\gamma$ radiation from Ws), so ideally a
 complete calculation as in the Double Pole Approximation of RacoonWW
 would be used.  As yet, this only produces weighted events so no
 realistic estimates are available.

 Hadronisation has been estimated by comparing models, \eg\ \Jetset\
 and \Herwig.  Other approaches include reweighting key variables in
 MC to correspond to data and propagating the effect to \Mw.  \Delphi\
 also advocate the use of so-called ``mixed Lorentz boosted \Zz''
 (MLBZ).

 The relative uncertainty in the \Lep\ beam energy enters directly
 into the uncertainty in \Mw, due to the use of kinematic fits which
 include a constraint to the centre-of-mass energy.  Uncertainties in
 beam energy are taken directly from the detailed studies of resonant
 depolarisation, NMR probes/flux loop measurements, and the \Lep\
 spectrometer project.

 A significant bias to the apparent W mass measured in the \WWqqqq\
 channel could arise if the hadronisation of the two W bosons is not
 independent and correctly modelled. Final state interactions such as
 CR and BEC may cause just this effect and are estimated by using
 phenomenological models. Direct searches for these effects, which may
 limit the viable set of such models, are described below

\section{Colour Reconnection}
 In \WWqqqq\ events, the products of the W decays in general have a
 significant space-time overlap as the separation of their decay
 vertices is small compared to characteristic hadronic distance
 scales.  Colour reconnection refers to a rearrangement of the colour
 flow between the two W bosons.  The effects of interactions between
 the colour singlets during the perturbative phase are expected to be
 small, $\sim (\frac{\as}{\pi N_{\mathnormal{colours}}})^2\Gamma_W$.
 The situation is less clear in the non-perturbative phase, where
 phenomenological models are implemented in hadronic Monte Carlos.  A
 higher susceptibility to CR (and more \Zqq\ background) is expected
 when \Wp\ and \Wm\ hadronisation regions overlap, so the spacetime
 picture of the QCD shower development is important.

 The predicted (barely) observable effects of CR include changes to the
 charged particle multiplicity, momentum distributions and the
  particle flow relative to the 4-jet topology.  The aim
  is to establish whether CR actually exists as well as
 controlling (or better, calibrating using data) the bias on \Mw\
 measurements.  The basic analysis method consists of comparing
 fully hadronic events with either: models with/without CR;
 \WWqqlv\ events; or MLBZs.

\subsection{Charged Particle Multiplicity}
\begin{table}[htbp]
 \begin{center}
 \begin{tabular}{|l|c|c|c|} \hline
    Expt.  &    \nQQQQ\     &    \nQQLV\      &  \Dnch\       \\ \hline
 \Aleph\  (183--202~GeV) \cite{bib:fsi_aleph}
& $35.75\pm0.54$ & $17.41\pm0.19$ & $+0.98\pm0.43$ \\
 \Delphi\ (183~GeV)  \cite{bib:fsi_delphi}
                 & $38.11\pm0.72$ & $19.78\pm0.65$ & $\nQQQQ/2\nQQLV$ \\
 \Delphi\ (189~GeV)   \cite{bib:fsi_delphi}
                 & $39.12\pm0.49$ & $19.49\pm0.41$ & $=0.981\pm0.027$\\
 \Lthree\ (183--202~GeV)  \cite{bib:fsi_l3}
                 & $37.90\pm0.43$ & $19.09\pm0.24$ & $-0.29\pm0.40$  \\
 \Opal\   (183~GeV)  \cite{bib:fsi_opal}
                 & $39.4\pm0.8$   & $19.3\pm0.4$   & $+0.7\pm1.0$   \\
 \Opal\   (189~GeV)  \cite{bib:fsi_opal}
                 & $38.31\pm0.44$ & $19.23\pm0.27$ & $-0.15\pm0.58$ \\
 \hline
 \end{tabular}
 \end{center}
 \caption{Summary of recent charged particle multiplicity measurements,
 with combined statistical and systematic errors. \Aleph\ measurements
 are not corrected for detector effects.}
 \label{tab:nch}
\end{table}

 The difference in multiplicities between \WWqqqq\ and twice the
 hadronic component of \WWqqlv\ events, $\Dnch=\nQQQQ-2\nQQLV$, should
 be zero in the absence of CR effects.  A compilation of the most
 recent multiplicity measurements is given in Table~\ref{tab:nch}.
 The effects are expected to be enhanced for softer particles, $p
 \lesssim \Gw$, so observables with implicit scale such as
 $\ln(1/x_p)$, $p_T$, rapidity, have also been studied with similarly
 inconclusive results.  As the systematics are now comparable to the
 statistical uncertainties, no attempt is made to combine results in a
 trivial way.  \Delphi\ and \Opal\ have studied the effects of CR for
 heavier hadron species, such as kaons and protons.  Although the
 numerical effects of CR are larger, the significance is reduced due
 to loss in statistical precision.

\subsection{Interjet Analysis}
\begin{figure}[htbp]
 \begin{minipage}{0.4\textwidth}
  \centerline{\mbox{\epsfig{file=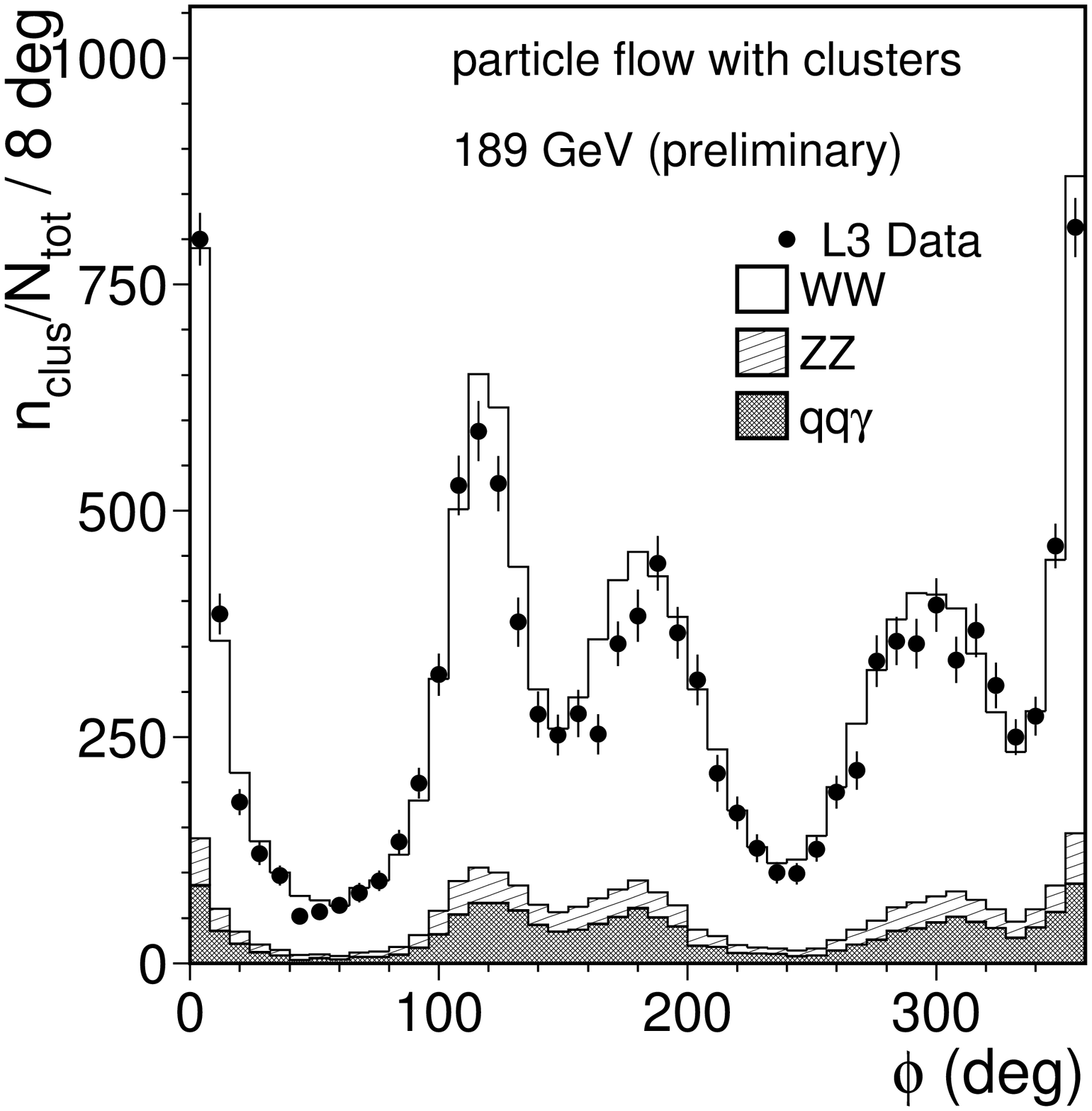,width=\textwidth}}}
  \caption{Particle density in a single di-jet plane.}
  \label{fig:pflow_uncorr}
 \end{minipage}
%
\hspace*{16mm}
 \begin{minipage}{0.4\textwidth}
  \centerline{\mbox{\epsfig{file=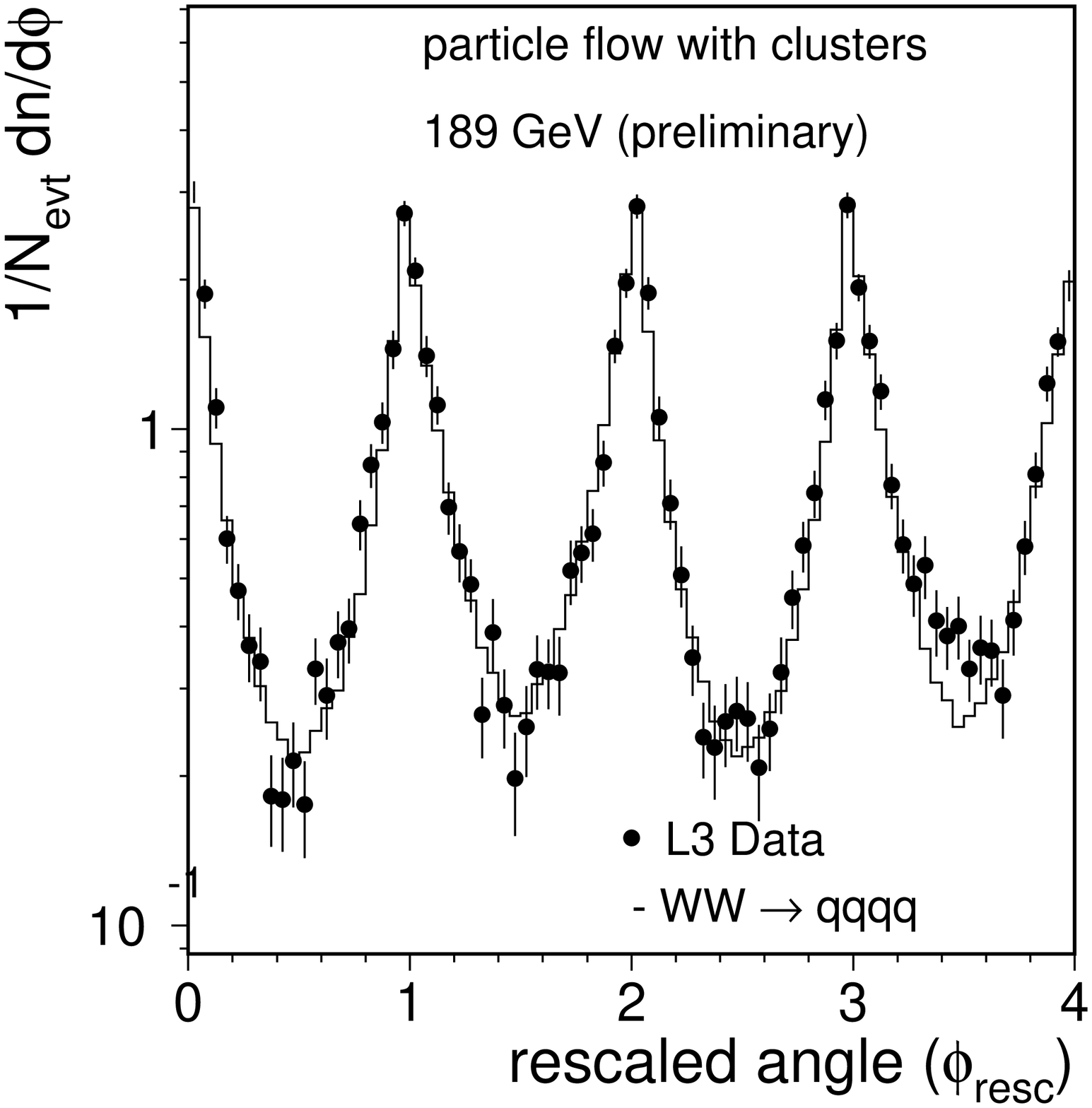,width=\textwidth}}}
  \caption{Particle density for all planes,
           scaled to di-jet opening angles.}
  \label{fig:pflow_corr}
 \end{minipage}
\vspace*{-5mm}
\end{figure}

  These analyses \cite{bib:fsi_aleph}, \cite{bib:fsi_l3}, \cite{bib:fsi_opal},
   are based on the ``string effect'' analysis in
  $\epem\rightarrow q\overline{q}g$ at lower energies.  All \Lep\
  collaborations are now using these analyses, recently developed by
  \Lthree\ and following earlier studies \cite{bib:interjets}. Events
  are selected by requiring, in addition to usual \WWqqqq\ criteria:
  four distinct jets ($y_{34}>0.01$ in \kt\ scheme); the two largest
  jet-jet angles are $100^\circ$--$140^\circ$ (intra-W regions); the
  two smallest jet-jet angles are $<100^\circ$ (inter-W regions) and
  have no jets in common.  This leads to an efficiency $\sim
  15$\%, good jet-jet to W association, with ``strings'' that are
  back-to-back and not crossing one another in 87\% of events.  In the
  case of \Lthree\ at $\roots=189$~GeV, 209 such events are selected.
  Next, particles are projected onto each of the four selected di-jet
  planes, forming the particle density as a function of angle from one
  of the plane defining jets.  This is illustrated in
  Figure~\ref{fig:pflow_uncorr}, where the inter-W regions correspond
  to the larger regions between the jet peaks.  As the di-jet angles
  vary from event to event and plane to plane, particle densities are
  scaled to the di-jet opening angle event-by-event, as show in
  Figure~\ref{fig:pflow_corr}, after background subtraction.
 \begin{figure}[h]
 \begin{minipage}[h]{0.49\textwidth}
 \centerline{\epsfig{file=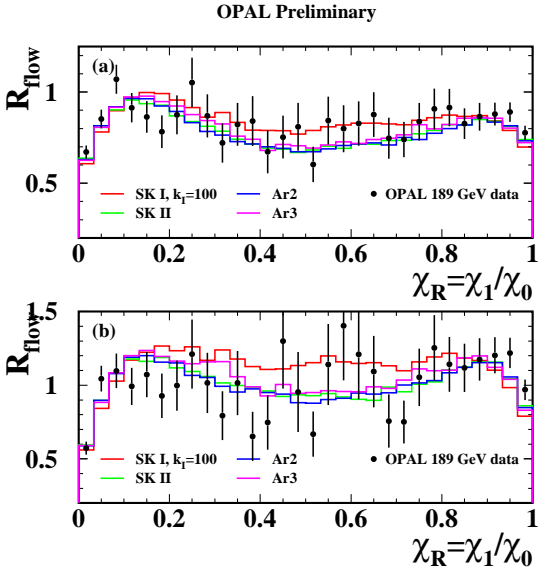,width=\textwidth}}
 \caption{Ratio of inter-W/intra-W particle flow (a) higher
          efficiency analysis, weighted by $\ln\frac{1}{x_p}$,
          (b) L3-like selection, see \protect\cite{bib:fsi_opal}
          for details.}
  \label{fig:pflow_ratio}
 \end{minipage}
 \hspace*{4mm}
 \begin{minipage}[h]{0.47\textwidth}
  To quantify the effects, the ratio of the inter-W/intra-W regions
  (or its inverse) is formed, as shown in Figure~\ref{fig:pflow_ratio},
  and integrated in the central region away from the jet peaks. An
  alternative is to consider the ratio of the integrals of inter-W and
  intra-W particle densities.  L3 estimate their sensitivity to the SK
  I CR model \cite{bib:SK}, is 3.2 (0.5) $\sigma$ total error for 100
  (32) \% reconnected events, using $\frac{1}{3}$ of their \Leptwo\ sample.
  \Opal\ have a variant on the analysis using a likelihood based
  selection to associate di-jets with Ws, a 4-C kinematic fit to
  define jet axes, and no double counting of particles, which predicts
  a slight improvement in sensitivity and favours $\sim 65$\% of
  events reconnected in the SK I model.  However, their emulation of
  the \Lthree\ analysis prefers a no-CR scenario.
 \end{minipage}
 \end{figure}

\subsection{Towards Mass Biases}
\begin{figure}[h]
\begin{minipage}{0.4\textwidth}
\begin{center}
  \mbox{\epsfig{file=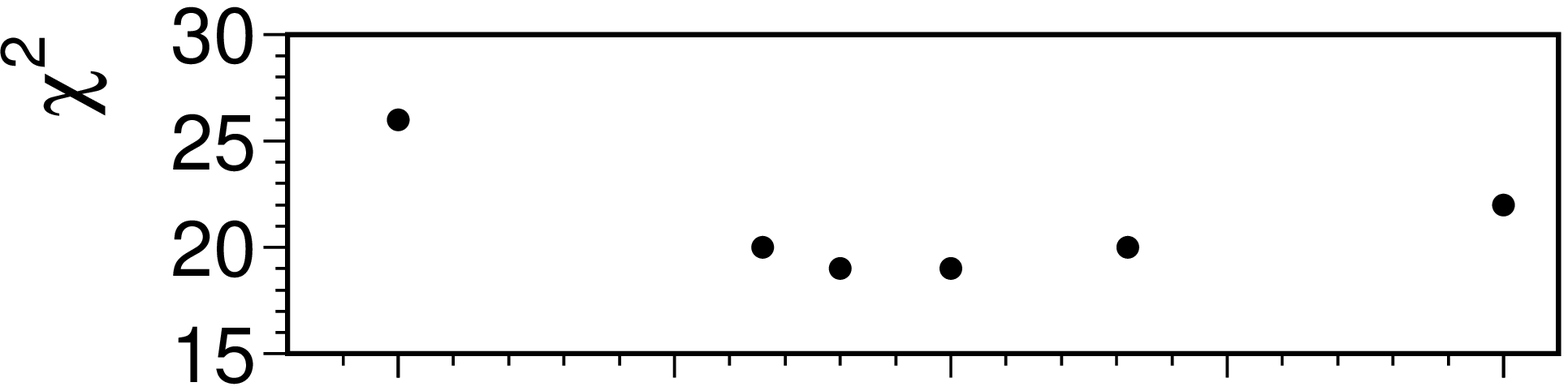,width=\textwidth}}
  \mbox{\epsfig{file=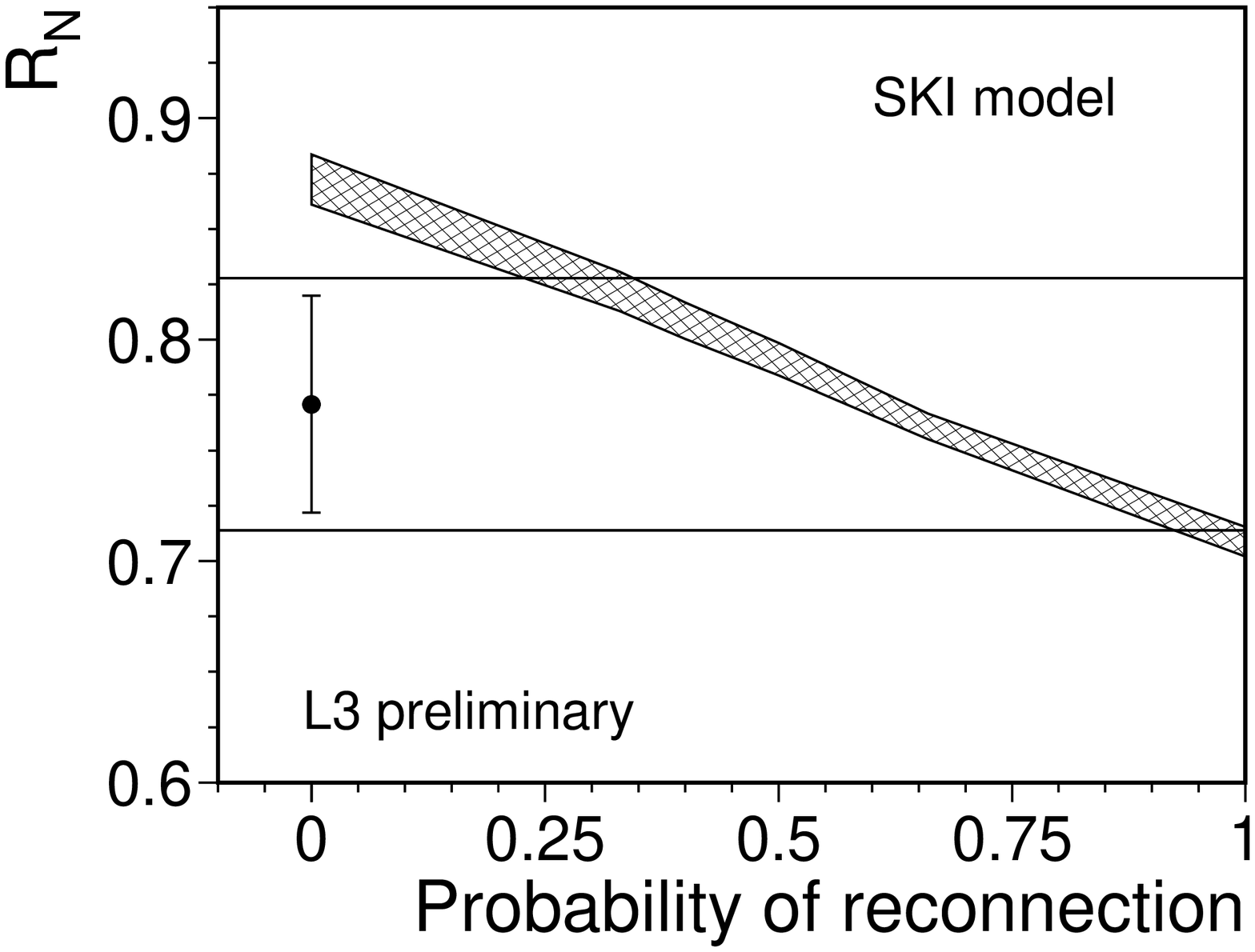,width=\textwidth}}
\end{center}
\end{minipage}
%
\hspace*{16mm}
\begin{minipage}{0.4\textwidth}
  \mbox{\epsfig{file=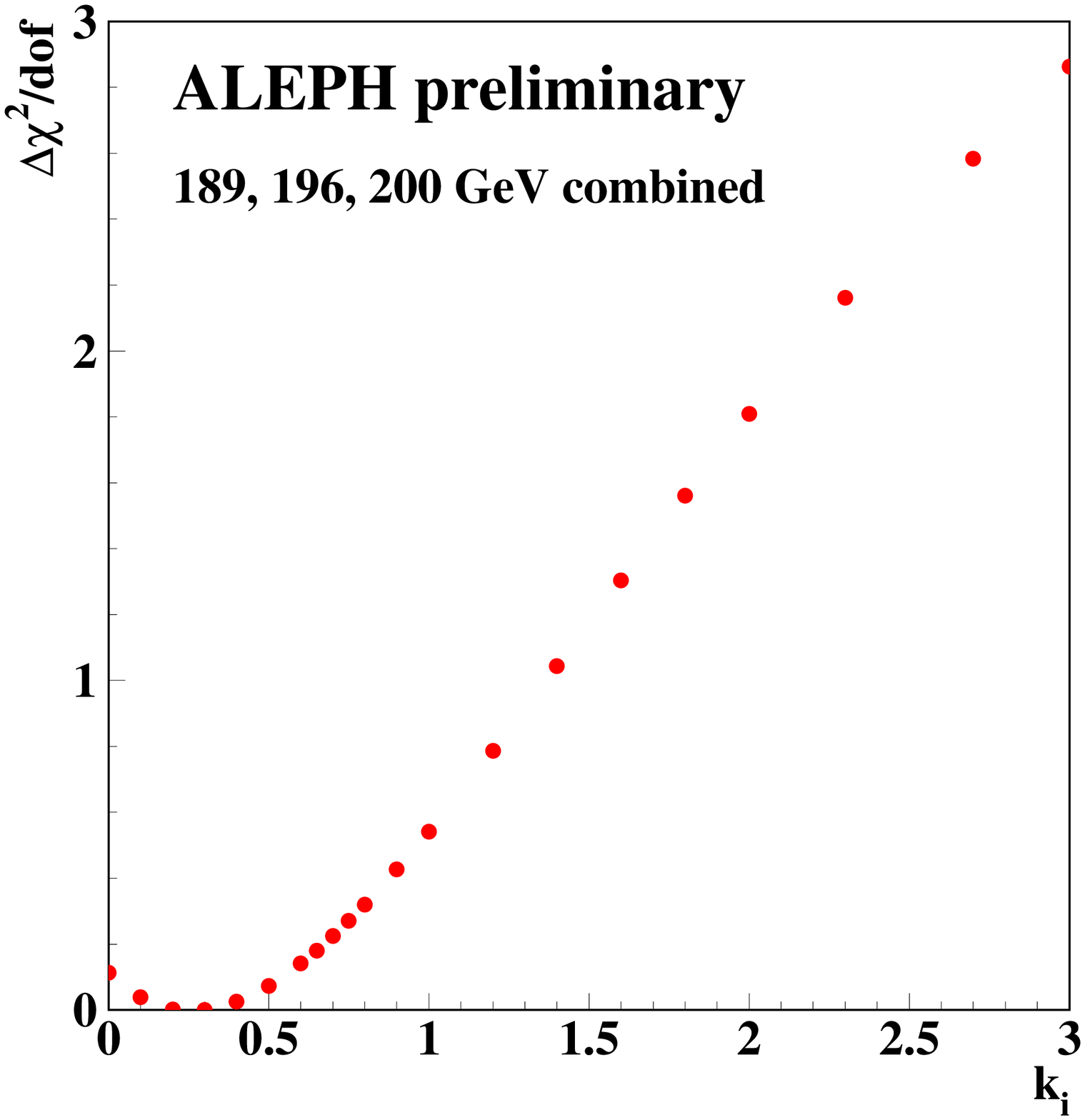,width=\textwidth}}
\end{minipage}
\caption{$\chi^2$ between data and SK CR model as a function of
reconnection probability.}
\label{fig:cr_data_aleph_l3}
\end{figure}
  In the SK I model, the reconnection probability is a free parameter
  and therefore can be adjusted to provide the best agreement of data.
  \Lthree\ estimate their data prefer $\sim40$\% CR in this model, and
  are 1.7$\sigma$ separated from no-CR scenario, as shown in
  Figure~\ref{fig:cr_data_aleph_l3}.  A similar study in \Aleph\
  suggests data prefer this model dependent probability to be
  $\sim15$\%, also shown in Figure~\ref{fig:cr_data_aleph_l3}.  Having
  estimated the preferred CR probability, it is trivial to estimate
  the corresponding bias on \Mw\ in the context of a single model.
  Comparison of such results between collaborations is complicated by
  differences in the models they use, such as hadronisation tuning.
  Using common samples of simulated events which are processed by each
  collaboration's detector simulation is in progress and will lead to
  a significant improvement in understanding in this area.

\section{Bose-Einstein Correlations}
Bose-Einstein correlation leads to the enhanced production of
identical boson pairs, such as \pipip\ or \pipim, at small 4-momentum
difference, $Q^2=-(\pb{1}-\pb{2})^2$.  This phenomena is firmly
established in various environments, in \Zqq\ at \Lepone\ and between
the particles of a single W boson at \Leptwo, among others.
Traditionally, it is studied using a 2-particle correlation function:
$R(\pb{1}, \pb{2})=\rho_2(\pb{1},\pb{2})/\rho_0(\pb{1}, \pb{2})$,
where $\rho_2$ and $\rho_0$ are 2-particle densities with and without
BEC, respectively.  Of particular interest in the \WW\ case is whether
or not there is any additional correlation introduced between the
decay products of the two Ws, which could potentially bias \Mw\
measurements.

One serious problem in this area is the construction of the reference
sample, $\rho_0$, and there are three frequently used methods.  The
first takes unlike-sign particle species, such as \pipid, forms the
ratio of like-sign/unlike-sign, and then takes the ratio of this
quantity relative to Monte Carlo to reduce the impact of having
resonances in $\rho_0$ but not in $\rho_2$ (``double ratio'').  The
second consists of taking $\rho_0$ to be a like-sign MC sample,
although this is subject to deficiencies in modelling.  The third
involves mixing pairs of data events, such as \Zqq\ or the
hadronically decaying W in \WWqqlv.  Another serious problem is the
ignorance of non-perturbative QCD amplitudes, forcing analyses to
resort to MC models, which in turn suffer from being probabilistic in
nature, with the models of BEC implemented in Monte Carlo
simulations.

A common parametrisation of the correlation function is
$R(Q)\sim1+\lambda\exp(-r^2Q^2)$, where $\lambda$ and $r$ represent
the source strength and size, respectively. Results from each of
\Aleph\ \cite{bib:fsi_aleph}, \Lthree\ \cite{bib:fsi_l3} and \Opal\
\cite{bib:fsi_opal} are summarised below. No \Delphi\ results were
presented (at their request), as they were undergoing substantial
revision.  Results which differed qualitatively from their earlier
analyses were presented at Moriond QCD 2001.

In the \Opal\ analysis, the two-particle
correlation function is formed using the double ratio of
like-sign/unlike-sign, data/no-BEC MC.  Three event classes are
identified, \WWqqqq\, \WWqqlv\ and \Zqq, each consisting of the linear
sum of pure contributions weighted by a probability derived from MC.
The classes are $C^{\mathnormal{DIFF}}$ (inter-W BEC),
$C^{\mathnormal{SAME}}$ (intra-W BEC), $C^{Z^*}$ (non-radiative
$q\overline{q}$. A simultaneous fit is performed to extract
$\lambda^{\mathnormal{DIFF}}$ for various source size hypotheses.
Examples are: $R^{\mathnormal{DIFF}}=R^{\mathnormal{SAME}}=R^{Z^*}$,
yielding $\lambda^{\mathnormal{DIFF}}=-0.14\pm0.36$, and completely
independent $R$s, giving $\lambda^{\mathnormal{DIFF}}=2.9\pm1.7$ and
$\lambda^{\mathnormal{SAME}}=0.62\pm0.10$. Intra-W BEC is
established, but the analysis is unable to ascertain whether inter-W
BEC exist.

In the \Aleph\ analysis, again the double ratio
is used, and the \Zqq\ background, in which BEC are known to be
present, is modelled using the model $BE_3$. It is concluded that
inter-W BEC is disfavoured by data with a significance of 2.2$\sigma$,
while the data are compatible with intra-W BEC.  An event mixing
analysis is also used, which qualitatively disfavours inter-W BEC, but
is as yet incomplete in that a full study of systematic errors has not
been performed.

The \Lthree\ analysis uses the method of Chekanov
 \etal\cite{bib:chekanov}, which sets out a very robust framework to
 test for the presence of inter-W BEC.  If the \Wp\ and \Wm\ decays
 are uncorrelated, then:
\begin{equation}
\rho_2^{\WW}(\pb{1},\pb{2})=\rho_2^{\Wp}(\pb{1},\pb{2}) +
                            \rho_2^{\Wm}(\pb{1},\pb{2}) +
                           2\rho_1^{\Wp}(\pb{1})\rho_1^{\Wm}(\pb{2}),
\label{eq:chekanov1}
\end{equation}
where the first two terms on the r.h.s. are estimated from individual
\WWqqlv\ events, assuming they are the same in \Wp\ and \Wm\ decays,
while the rightmost term is formed by mixing pairs of \WWqqlv\ events
to give ``fake'' events in which there can be no true inter-W BEC.
All background samples include BEC using the BE$_{32}$ model.
 \begin{figure}[h]
 \begin{minipage}[h]{0.49\textwidth}
  \centerline{\epsfig{file=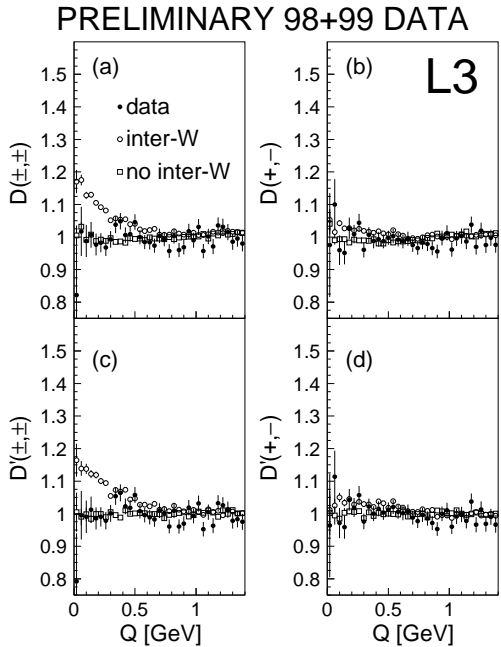,width=\textwidth}}
  \caption{Inter-W BEC observables, see text for details.}
  \label{fig:bec_l3}
 \end{minipage}
 \begin{minipage}[h]{0.49\textwidth}
 The variable $D$ is defined as the ratio of the left hand side to the
right hand side of Equation~\ref{eq:chekanov1}.  In the absence of
inter-W correlations, such as from BEC, and of bias introduced in the
event mixing, $D=1$.  To eliminate this potential residual
experimental bias, the variable $D^{\prime}$ is constructed as ratio
of $D$ in data to that in MC (having only intra-W BEC).
Figures~\ref{fig:bec_l3}(a)-(b) and ~\ref{fig:bec_l3}(c)-(d) show the
variables $D$ for like-sign and unlike-sign data, and similarly for
$D^{\prime}$.  The data clearly favour the no inter-W BEC model.
By fitting the phenomenological correlation function to these data,
\Lthree\ obtain $\lambda=0.013\pm0.018\pm0.015$, where $\lambda=0$
corresponds to no inter-W BEC. In contrast, the BE$_{32}$ model gives
$\lambda=0.126\pm0.006 \mathnormal{(stat.)}$, thus the data disfavour
inter-W BEC by 4.7$\sigma$.
\end{minipage}
\end{figure}

\section{Combined \Lep\ Results and Summary}
\begin{figure}[htb]
\begin{minipage}{0.4\textwidth}
\begin{center}
  \mbox{\epsfig{file=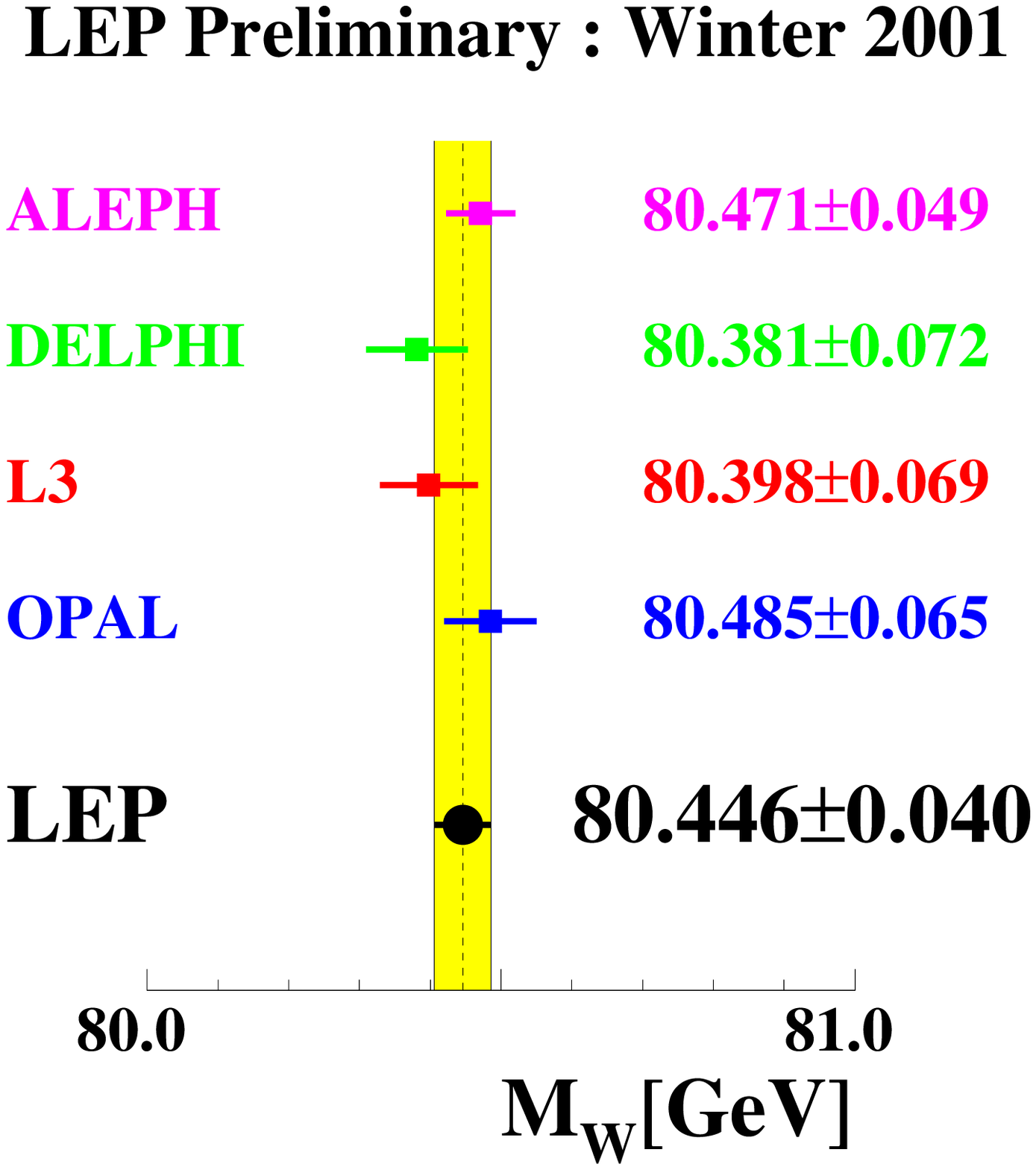,width=\textwidth}}
\end{center}
\end{minipage}
\hspace*{12mm}
\begin{minipage}{0.4\textwidth}
\begin{center}
  \mbox{\epsfig{file=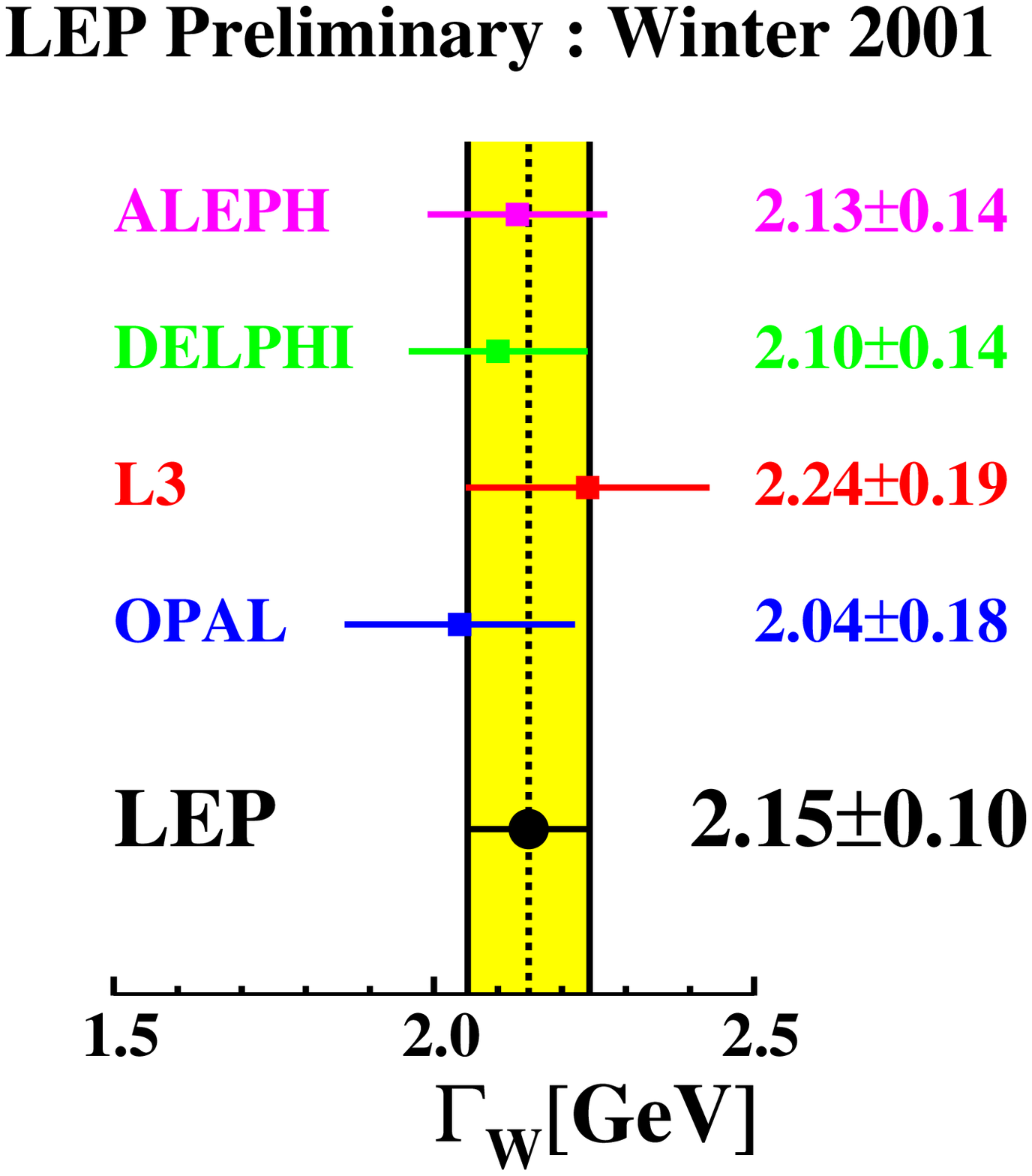,width=\textwidth}}
\end{center}
\end{minipage}
\caption{Combined \Lep\ Mw and \Gw, all channels.}
\label{fig:lepmwgw}
\end{figure}

\begin{figure}[htb]
\begin{minipage}{0.4\textwidth}
\begin{center}
  \mbox{\epsfig{file=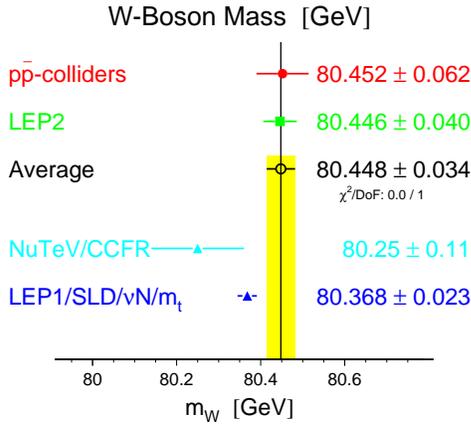,width=\textwidth}}
\end{center}
\end{minipage}
\hspace*{12mm}
\begin{minipage}{0.4\textwidth}
\begin{center}
  \mbox{\epsfig{file=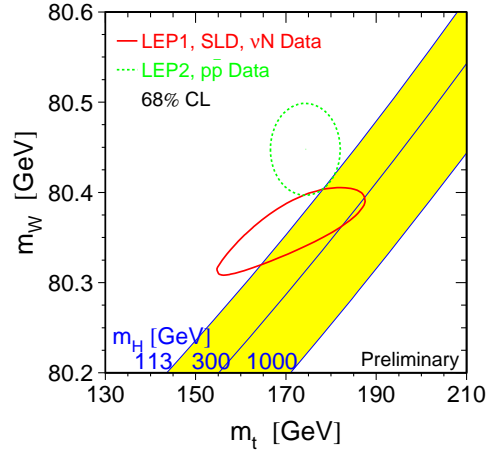,width=\textwidth}}
\end{center}
\end{minipage}
\caption{Comparison of \Lep\ \Mw\ with other measurements and Standard
Model predictions}
\label{fig:lep_interpretation}
\end{figure}
Based on an analysis of 2.8~fb$^{-1}$ ($\sim82$\% of the entire
\Leptwo\ data sample, 100\% analysed by \Aleph\ and \Lthree), the
following preliminary measurements \cite{bib:lepmw_combined},
summarised in Figures~\ref{fig:lepmwgw} and
~\ref{fig:lep_interpretation}, are made: $\Mw =
80.446\pm0.026\pm0.030$~GeV, $\Gw = 2.148\pm0.071\pm0.063$~GeV,
$\Delta_{\Mw}(\qqqq-\qqlv) = +18\pm46$~MeV.  Further improvements in
systematics are anticipated.  Colour reconnection analyses, such as
interjet multiplicity, from all collaborations will be combined.  The
overall conclusion regarding inter-W Bose-Einstein correlations among
the \Lep\ collaborations is finally becoming more consistent: there is
increasingly strong evidence that they do not exist.

\section*{Acknowledgements}
Work supported by PPARC grant GR/L04207.  I would like to thank
colleagues in all \Lep\ collaborations for the preparation and timely
combination of results presented.

\end{document}